\documentclass[lettersize,journal]{IEEEtran}
\usepackage{amsmath,amsfonts}
\usepackage{algorithmic}
\usepackage{algorithm}
\usepackage{array}
\usepackage{textcomp}
\usepackage{stfloats}
\usepackage{url}
\usepackage{xcolor}
\usepackage{verbatim}
\usepackage{graphicx}
\usepackage{cite}
\usepackage{subcaption}
\usepackage{caption}
\usepackage{amsmath}
\usepackage{multirow}
\usepackage{booktabs}
\usepackage{adjustbox}
\usepackage{makecell}
\begin{document}

\title{Cyber Resilience Assessment of Unbalanced Distribution System Restoration under Sparse Load Forecasting Attacks}

\author{Chen Chao,~\IEEEmembership{Graduate Student Member,~IEEE};
Zixiao Ma,~\IEEEmembership{Member,~IEEE};
Ziang Zhang,~\IEEEmembership{Senior Member,~IEEE}

\thanks{The authors are with the Department of Electrical and Computer Engineering, Binghamton University, State University of New York, Binghamton, NY 13902 USA (e-mail: cchao3@binghamton.edu; zma10@binghamton.edu; ziang.zhang@binghamton.edu).}
}



\maketitle

\begin{abstract}
System restoration is critical for power-system resilience, but its growing reliance on artificial intelligence (AI)-based load forecasting creates a cyber-physical vulnerability in the restoration decision loop. 
Manipulated forecasts can cause infeasible restoration schedules, insufficient inverter-based-resource ramping margins, and unsuccessful recovery of de-energized segments, yet the resilience of restoration processes to such attacks remains largely unexplored.
This paper evaluates restoration vulnerability at the system level rather than only measuring forecasting error. A gradient-based sparse perturbation method is developed as a stress-testing tool to identify influential forecasting inputs.
We further create a restoration-aware validation framework that embeds these compromised forecasts into a sequential restoration model and evaluates operational feasibility using an unbalanced three-phase optimal power flow formulation. 
Case studies on a modified IEEE 123-bus feeder show that sparse input perturbations can substantially increase forecasting error and make selected microgrid restoration stages infeasible.
The results reveal system-level failures caused by active-power-balance infeasibility and power ramping violations, which can prevent the restoration of critical loads. 
These findings provide actionable insights for designing cybersecurity-aware restoration planning frameworks.
\end{abstract}


\section{Introduction}
\IEEEPARstart{T}he U.S. Department of Energy recently warned that by 2030, the number of power outages in the United States could increase by up to 100 times. This alarm is driven by aging infrastructure, rising electricity demand, rapid distributed energy resource (DER) growth, increasing cyber-physical complexity, and the surge of artificial intelligence (AI)‑driven data centers with challenging load profiles \cite{doe2025}. As large-scale outages become more frequent, the ability to restore power quickly and reliably, especially at the distribution level, becomes vital to grid resilience and public safety.

Distribution system restoration refers to re-energizing network segments after a blackout, often through localized microgrid (MG) formation and DER dispatch\cite{chen2017modernizing}. While extensive research has addressed restoration strategies such as optimization-based scheduling \cite{zhang2021two, maharjan2025distribution} and reinforcement learning approach \cite{vu2023multi}\cite{fazlhashemi2025distribution}, few have examined the growing risk posed by the shift toward data-driven restoration planning and its associated cyber vulnerabilities. This vulnerability stems from a critical dependency.
Modern restoration increasingly relies on predicted load information generated by AI-based forecasting models\cite{epri_ps39a_2025}. 
These models rely on external and often unverified inputs such as weather data which lacks rigorous cryptographic authentication \cite{nist_ot_2023}.
Thus they present a new cyberattack surface that remains largely unaddressed in current restoration frameworks.
Such vulnerabilities can be exploited by attackers with the objective of disrupting restoration-plan feasibility and preventing critical load pickup.

To better understand how this vulnerability emerged, it is important to review how restoration planning has historically addressed load uncertainty. Earlier approaches relied on cold load pickup (CLPU) modeling and stochastic or robust optimization frameworks \cite{xie2023dynamic} \cite{song2020robust}, aiming to capture demand variability and surge behavior following blackouts. However, accurately modeling post-blackout load behavior remains highly challenging due to human-driven patterns, DER interactions, and temporal non-stationarity. To overcome these challenges, recent efforts have shifted toward integrating data-driven short-term load forecasting into the restoration process.
For example, forecasted load profiles are integrated into the optimization model for post-blackout distribution system restoration in \cite{konar2023mpc}. At the transmission level, real-time measurements of the restored load blocks are used to update statistical forecasts for unrecovered areas in \cite{edib2023situation}.
Forecast-based restoration planning offers improved adaptability and feasibility by aligning load estimations with real-world recovery dynamics.

To improve load estimation in restoration contexts, many recent studies have adopted AI models for short-term load forecasting. These models offer strong nonlinear modeling capabilities and have shown superior performance over traditional techniques. For example, long short-term memory (LSTM) networks \cite{kong2017short}, hybrid Convolutional Neural Network (CNN)-LSTM architectures \cite{tian2018deep}, and transformer-based models \cite{fan2024optimizing} have demonstrated improved accuracy in capturing complex temporal dependencies and multivariate patterns, particularly under high variability conditions.
However, their growing complexity also increases models' sensitivity to data quality. Most AI-based forecasters rely on external inputs such as weather and behavioral data, which may be unverified or easily manipulated \cite{zhang2024vulnerability, liu2024enhancing}. As a result, the forecasting layer can become an entry point for cyber-physical disruptions in restoration planning.
Recent research has shown that even imperceptible input perturbations can significantly degrade forecasting accuracy, including under black-box settings. In \cite{chen2018machine}, adversarially crafted weather data led to large forecasting errors, and \cite{chen2019exploiting} showed that spoofed public-weather inputs caused costly dispatch mistakes and unintended load shedding. To defend against such threats, countermeasures have been proposed, such as Bayesian adversarial training \cite{zhou2022robust} and AI-based anomaly detection \cite{cui2019machine}.
These studies demonstrate forecasting-layer vulnerability, but the downstream impacts of such attacks, particularly their potential to undermine restoration feasibility and trigger unsafe grid conditions, have received limited attention.

Despite growing efforts to secure AI models, most existing research focuses on applying forecasting loads to normal grid operations.
Little attention has been paid to how adversarial forecasting errors impact grid behavior during restoration, when the system is at its most vulnerable. In particular, the low-inertia nature of inverter-dominated microgrids and the constrained balancing capabilities during black-start phases make restoration highly sensitive to load mismatches.
Thus, one central restoration question is not only whether an attacker can increase forecasting error, but whether that error can violate operating constraints and leave critical loads unrestored.
This paper addresses this overlooked cyber-physical vulnerability by developing an attack-restoration-validation framework that quantifies how compromised load forecasts degrade restoration feasibility and operational security.
Specifically, we first develop a cyber-resilience threatening method driven by a novel gradient-based sparse adversarial attack. 
Then, to assess its impact on downstream restoration planning, a three-phase unbalanced optimal power flow (OPF) framework is proposed to validate the operational feasibility of resulting plans under true load conditions.


The main contribution of this paper is threefold:
\begin{itemize}
\item 
This paper develops a cyber-physical vulnerability assessment framework for AI-based forecast-dependent distribution system restoration. The framework traces compromised forecasts through restoration scheduling and feasibility validation under actual post-outage load conditions.
\item 
We propose a gradient-based stress-testing method to identify high-impact weather-input entries and generate sparse load-forecast perturbations. It is extended to a query-based black-box setting without requiring forecasting-model parameters.
\item 
A system-level validation framework is developed by integrating CLPU-based restoration planning and three-phase unbalanced OPF, with a restoration impact index introduced to quantify infeasible restoration stages, IBR ramping violations, and failed critical-load pickup.
\end{itemize}

\section{Problem Statement}
\begin{figure*}
    \centering
    \includegraphics[trim=30 15 15 620, clip ,width=1.0\linewidth]{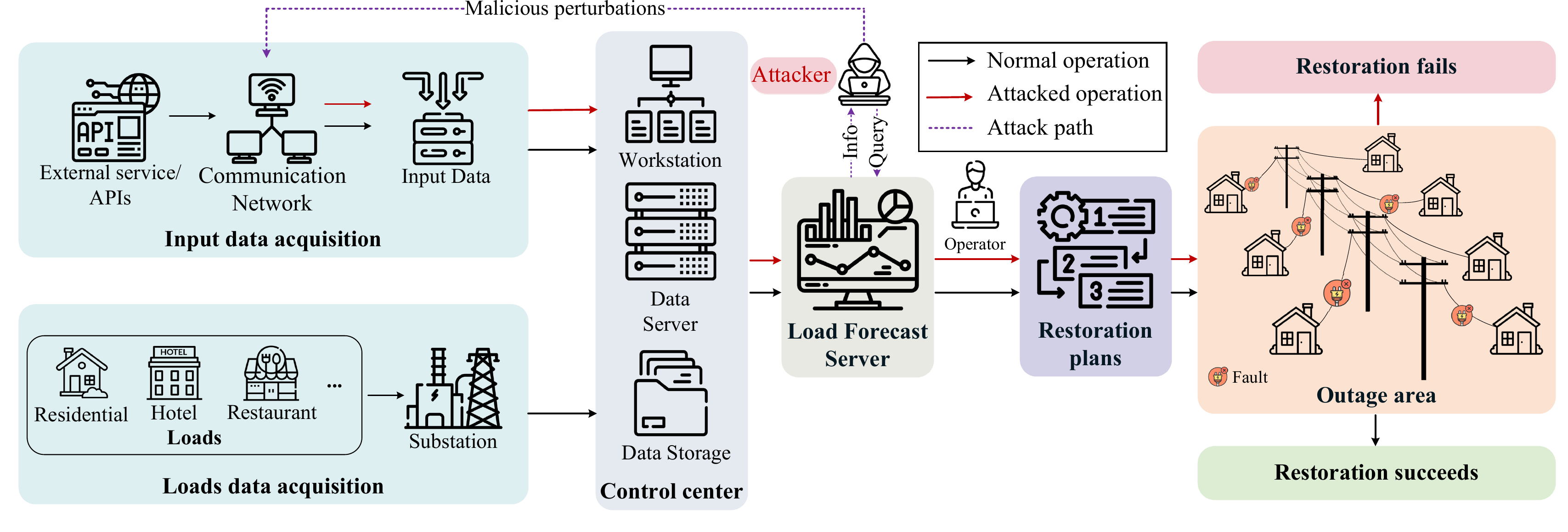}
    \caption{
    Illustration of how load-forecasting attacks can propagate into distribution system restoration. Some external inputs are modified by attackers and processed by the load-forecasting server. System operators then generate restoration plans from compromised forecasts, which can lead to infeasible load pickup and restoration failure under actual load conditions.
    }
    \label{fig:overall figure}
\end{figure*}
The increasing integration of load forecasting into distribution system restoration introduces a cyber-physical attack surface in which data perturbations can affect load-pickup decisions feasibility.
This section explains how forecasted loads enter restoration decisions and formulates adversarial forecast tampering as a mechanism for stressing restoration feasibility.

\subsection{Role of Load Forecasting in Restoration and Its Vulnerabilities to Cyber Attacks}
After a blackout, the restoration process is typically divided into three main phases: preparation, system restoration, and load restoration \cite{liu2016power}. 
As an offline decision-making part of this process, restoration planning relies on forecasting loads to determine recovery sequences.
Generally, it can be formulated as an optimization model, where load forecasting results are often embedded in the objective function and key constraints like nodal power balance.
If these results are perturbed, the restoration plan will be scheduled on incorrect demand predictions.
Execution of such a plan can lead to generation-load mismatch, active-power-balance infeasibility, and unsuccessful load pickup.  
The critical dependency thus creates a potential cyber attack surface.

Fig. \ref{fig:overall figure} illustrates this vulnerability across the restoration process, where an attacker seeks to induce forecast deviations that render restoration strategies ineffective under actual system conditions.
In the data acquisition stage, some input data are obtained from external services or APIs through communication networks, while load data are collected at substations and transmitted to the control center for storage and processing.
After that, these data are sent to load forecasting server and system operators use forecasting results to make restoration plans for outage areas.
For attackers, they can query the forecasting server and use the responses to craft malicious perturbations on input data through the communication network. This then leads to serial attacked operations and finally restoration failing in outage areas.
Therefore, it is imperative to investigate the vulnerability of power system restoration under adversarial load forecasting.

\subsection{Formulation of Adversarial Attack on Load Forecasting}

\subsubsection{Load forecasting formulation}
Typically, load forecasting models are trained on historical load data, time indicators, and weather-related influence factors.
The system operator would be able to collect a training data set \(D_{t r}=\left\{\left(\textbf{X}_{t-H}, \ldots, \textbf{X}_{t-1}\right) ; L_{t}\right\}_{t=H}^{T}\) where \(L_{t}\) is the load value at time t; \(\textbf{X}_{t-i}=\left(L_{t-i}, X_{t-i}^{w}, X_{t-i}^{\text {index }}\right)\) is the historical data at time \(t-i\) (\( 1\le i\le H \)) including the load value \(L_{t-i}\), weather data \(X_{t-i}^{w}\), and time index \(X_{t-i}^{\text {index }}\). 
Let $\textbf{X} = (\textbf{X}_{t-H}, \ldots, \textbf{X}_{t-1})^T$ be a $H \times J$ matrix where $H$ is the time length of influence factors and $J$ is the number of the input features.
The forecasting model can be formulated as a function parameterized by \(\theta: f_{\theta}\left(\textbf{X}\right)\).
Estimation of \(\theta\) is given by minimizing the \(L_2\)-norm of the difference between model predictions and ground truth loads:
\begin{equation}
    \begin{aligned}
\min _{\theta} & \text{ }\frac{1}{T-H+1} \sum_{t=1}^{T}\left\|f_{\theta}\left(\textbf{X}\right)-L_{t}\right\|_{2} \\
\text { s.t.  } & \text{ }\theta \in \Theta
\end{aligned}
\end{equation}
where \(T-H+1\) represents the number of samples in the training set. 



\subsubsection{Attacker's Objective on Load forecasting}
Adversarial attacks refer to attacks on machine learning (ML) algorithms that perturb the input data to manipulate the model’s prediction during testing stage \cite{chakraborty2021survey}. 
Under the load forecasting scenario, the attacker aims to deviate the forecasted loads from a reference value $L_t^{\mathrm{ref}}$, which can be estimated from historical profiles or a surrogate forecast. The attack is formulated as:
\begin{equation}
\label{eq:attack formulation}
\begin{aligned}
\max \textbf{ } &\mathcal{L} (f_{\theta}(\tilde{\textbf{X}}), L_t^{\mathrm{ref}} )\\
\text { s.t. } 
&\left\|\tilde{\textbf{X}}_{t-i, j}-\textbf{X}_{t-i, j}\right\| \leq \epsilon,  \forall i ,j
\end{aligned}
\end{equation}
where \(\tilde{\textbf{X}}\) is the perturbed testing input. 
In this work, weather inputs \(X_{t-i}^{w}\) are chosen as the attack target following most existing works since they are more vulnerable compared to other input features\cite{zhang2024vulnerability, liu2024enhancing, chen2018machine, chen2019exploiting}.
Unlike internal SCADA measurements, weather inputs are often obtained from external third-party APIs, so the exchanged data must go through external communication channels that expose them to interception or tampering \cite{nist_api_protection,owasp_tls}.
Meanwhile, as load forecasting models are increasingly deployed on cloud-based platforms, attackers can query black-box forecasting services and estimate their information without breaching internal networks \cite{ibrahim2022cloud}.

\subsubsection{Attacker's Knowledge} 
According to attackers' knowledge of the forecasting model, adversarial attacks can be divided into white box and black box scenarios. In white-box attacks, the attacker is assumed to have full information about the forecasting model including its structure and parameters. 
This enables the design of highly well-constructed perturbations. 
In contrast, black-box attacks operate under more constrained conditions
In this paper, we assume that attackers have query access to the load forecasting model but are blind to the forecasting algorithms and have no knowledge of any parameters of $f_\theta$.
The details of white box and black box cases are presented in Section $\mathrm{III}$. 

In summary, the increasing reliance of system restoration on ML-based load forecasting introduces a novel and critical vulnerability: by perturbing publicly sourced inputs such as weather data, an attacker can compromise restoration planning and system stability. In the next section, we develop concrete methods to construct such attacks.

\section{Designing Adversarial Load Forecasting Attacks for Power System Restoration} 

In this section, we develop a Sparse Adversarial Attack (SAA) as a diagnostic stress-testing tool for the forecasting layer of the restoration pipeline.
Rather than treating forecasting degradation as the final outcome, SAA identifies sparse input perturbations whose effects can be propagated through restoration planning and OPF validation.
Also, we extend the method to black-box settings using query-based finite difference gradient estimation.

\subsection{Sparse Adversarial Attack for Forecast Tampering}

Standard adversarial techniques for ML-based load forecasting, such as Projected Gradient Descent (PGD)\cite{kurakin2018adversarial}, can maximize model deviation but often perturb a predefined feature or a dense set of input entries. This demonstrates model vulnerability but does not identify which spatiotemporal input points are most responsible for downstream restoration infeasibility. Dense perturbations are also less stealthy and more likely to be detected by operators or anomaly detectors such as Bad Data Detection \cite{gu2013bad}.

To address this limitation, we propose a Sparse Adversarial Attack (SAA) that identifies the top vulnerable points to distort load forecasting results.
Based on gradient information, SAA picks the most sensitive spatiotemporal input coordinates and applies perturbations only at those locations. 
In this way, SAA serves not only as an attack method, but also as a targeted vulnerability assessment tool that reveals how small and stealthy perturbations can propagate through the load forecasting-restoration pipeline and trigger restoration failure.
The detail is presented in Algorithm 1.
In each iteration, after getting the gradient matrix $\textbf{g}$, we generate a binary mask matrix $\textbf{M}$ to help locate attack positions.
First, the top-$n$ elements are selected based on the absolute gradients and their positions in the input matrix are recorded in set $\mathcal{S}$:
\begin{equation}
    \mathcal{S} = [(i_1, j_1), \cdots, (i_n, j_n)]
\end{equation}
Then, entries in $\textbf{M}$ are defined based on the $\mathcal{S}$:
    \[
    \textbf{M}[i,j] = \begin{cases}
    1 & \text{if } (i,j) \in \mathcal{S} \\
    0 & \text{otherwise}
    \end{cases}, \forall i, j
    \]
Finally, the $\tilde{X}^{k}$ is updated using clip function: 
\begin{equation}
    \tilde{\textbf{X}}^{k} = \operatorname{Clip}_{\textbf{X}, \epsilon}\{\tilde{\textbf{X}}^{k-1} +\alpha \cdot \operatorname{sign}(\textbf{g}) \odot \textbf{M}\}
\end{equation}
where $\odot$ denotes the Hadamard product.

\begin{algorithm}[htbp]
\caption{Sparse Adversarial Perturbation attack}
\begin{algorithmic}
\STATE\hspace{-0.25cm} \textbf{Input:\\}
\hspace{0.25cm}$f_\theta$: forecasting model; $\textbf{X}$: input matrix;\\
\hspace{0.25cm} $L_t^{\mathrm{ref}}$: reference load; $\alpha$: step size; $\epsilon$: clipping bound; \\
\hspace{0.25cm}$K$: iterations; $n$: sparsity level
\STATE\hspace{-0.25cm} \textbf{Output:\\}
\STATE \vspace{0.1em}
\hspace{0.25cm}$\tilde{\textbf{X}}^{K}$: adversarial examples
\end{algorithmic}

\begin{algorithmic}[1]
\STATE Initialize $\tilde{\textbf{X}}^{0} \leftarrow \textbf{X}$ 

\FOR{$k = 1$ to $K$}
    
    \STATE Compute gradient: \( \textbf{g} \leftarrow \nabla_{\tilde{\textbf{X}}^{k-1}} \mathcal{L}(f_\theta(\tilde{\textbf{X}}^{k-1}), L_t^{\mathrm{ref}})\)
    
    \STATE Binary mask matrix: $\textbf{M} \in \mathbb{R}^{H \times J}$
    
    \STATE Identify top-$n$ elements: 
    \[ \text{Top-}n(\left| \mathbf{g}_{i, j} \right|) \text{ for } i \in [1, H], j = 2, \dots, 5\]

    \STATE Record top-$n$ elements' coordinate in $\tilde{\textbf{X}}^{k-1}$ to $\mathcal{S}$: 
    \[ \mathcal{S} = [(i_1, j_1), \cdots, (i_n, j_n)] \]
    
    \STATE Update mask matrix $\textbf{M}$:
    \[
    \textbf{M}[i,j] = \begin{cases}
    1 & \text{if } (i,j) \in \mathcal{S} \\
    0 & \text{otherwise}
    \end{cases}, \forall i, j
    \]

    \STATE Apply sparse update:
    \[
    \tilde{\textbf{X}}^{k} \leftarrow \operatorname{Clip}_{\textbf{X}, \epsilon}\{\tilde{\textbf{X}}^{k-1} +\alpha \cdot \operatorname{sign}(\textbf{g}) \odot \textbf{M}\}
    \]
\ENDFOR
\RETURN $\tilde{\textbf{X}}^{K}$
\end{algorithmic}
\end{algorithm}

\subsection{Adapting Attacks to Black-box Load Forecasting Models}

Black-box attacks are more practical, yet challenging, for load forecasting in power systems, since deployed models are often hosted on secure platforms without public access to their internal details.
In this context, the internal parameters or architectures of the ML models being attacked are not known. 
In such scenarios, the key to launching attacks is inferring gradient information, which characterizes the sensitivity of each input element to prediction errors. 
Perturbing elements with larger gradient magnitudes is more influential in deviating the model's prediction.
In this way, if the attacker is able to estimate gradients by querying the load forecasting model, black-box attacks remain feasible even in constrained settings. 

To be specific, for the $j$-th feature at time $t-i$, the attacker needs to query the load forecasting system to calculate the two-sided estimation:
\begin{equation}
    \nabla_{\tilde{\mathbf{X}}_{t-i, j}} \mathcal{L} \approx \frac{ \mathcal{L}(f_{\theta}(\tilde{\mathbf{X}}+\delta \mathbf{e}_{i,j}), L_t^{\mathrm{ref}})-\mathcal{L}(f_{\theta}(\tilde{\mathbf{X}}-\delta \mathbf{e}_{i,j}), L_t^{\mathrm{ref}})}{2 \delta}
\end{equation}
where $ \mathbf{e}_{i,j}$ is a $H\times J$ matrix with all zero except 1 at $(i, j)$, and $\delta$ is a small value for gradient approximation.
After this, we can follow the same steps in the above attack formulations to construct adversarial input.



\begin{figure}
    \centering
    \includegraphics[trim=15 15 100 425, clip ,width=1.0\linewidth]{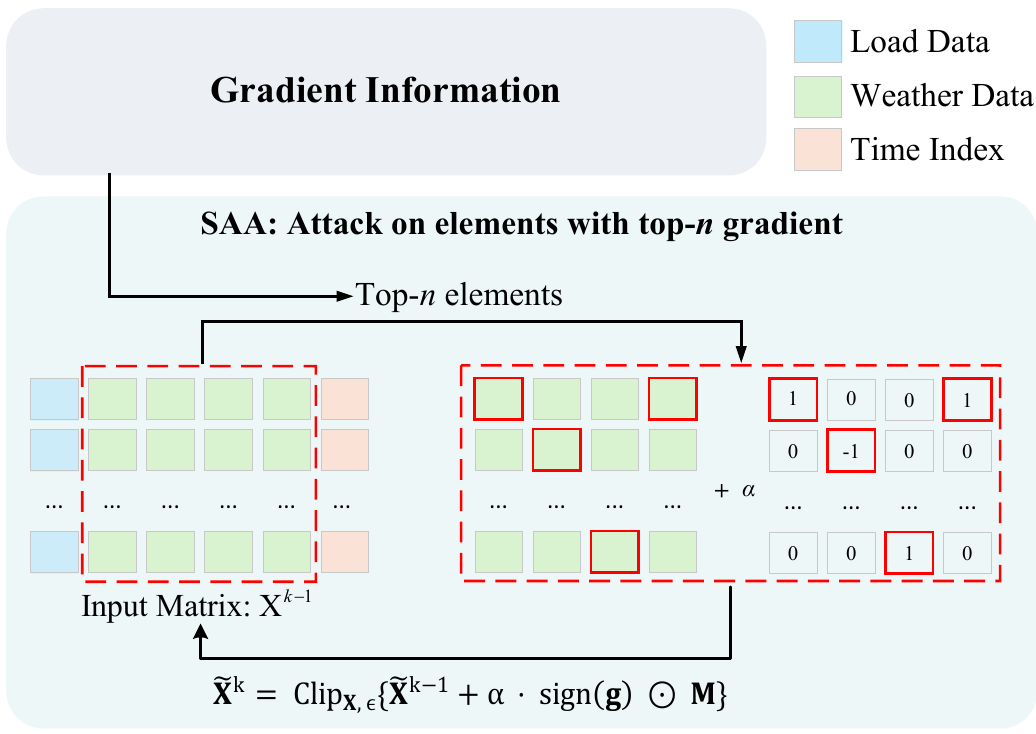}
    \caption{Illustration of sparse adversarial attack methods on load forecasting. It updates adversarial examples based on gradient information: picking top-$n$ elements within input matrix. In white box settings, gradient information can be directly obtained whereas under black box scenario, it can be estimated by querying the forecasting model.}
    \label{fig:attack comparison}
\end{figure}

In summary, SAA targets only influential input elements, reducing perturbation density while maintaining attack effectiveness. This sparse pattern supports restoration-oriented vulnerability assessment by identifying input coordinates most likely to affect downstream feasibility.
Fig. \ref{fig:attack comparison} illustrated how proposed SAA strategies modify input matrix.
In the next section, we will introduce how attacked forecasts impact restoration planning and how to evaluate the attack's practical impact.

\section{Power System Restoration under Adversarial Load Forecasting Attacks}
In this section, we connect the sparse forecast perturbations generated by SAA to the physical restoration process. 
We analyze vulnerabilities of restoration process under adversarial attacks, and to evaluate the impact of attacked restoration plans, we propose a three-phase unbalanced OPF validation framework and a Restoration Impact Index to quantify their downstream restoration consequences.

\subsection{Restoration Planning Model Based on Forecasted Loads }

Restoration planning is typically formulated as a mixed-integer linear programming (MILP) problem, where load forecasting results $f_\theta(\textbf{X})$ serve as key inputs to the restoration decisions.
In this work, these forecasting inputs are replaced by the attacked forecasts generated by SAA in Section III, so that the sparse vulnerable points identified in the forecasting stage can be propagated to their downstream influence on restoration scheduling and feasibility.
In the following, we use the MILP model for three-phase unbalanced distribution system restoration in \cite{zhang2021two} and present how load forecasting values are embedded in the restoration decisions.
{Note that though adopting a MILP formulation in this study, any restoration method that relies on load forecasts remains susceptible to adversarial perturbations of these forecasts.}

The model's objective is to maximize the total restored loads with priority factor $w_{i}^{\mathrm{L}}$ over the restoration period [1, T]:
\begin{equation}
\label{eq: restoration objective}
    \max \sum_{t \in[1, T]} \sum_{i \in \Omega_{L}} \sum_{\phi \in \Omega_{\phi}}\left(w_{i}^{\mathrm{L} }x_{i, t}^{\mathrm{L}} P_{i, \phi, t}^{\mathrm{L}}\right)
\end{equation}
where  $P_{i, \phi, t}^{\mathrm{L}}$ and $x_{i, t}^{\mathrm{L}}$ are the restored load and its restoration status at time $t$. If $P_{i, \phi, t}^{\mathrm{L}}$ is restored, then $x_{i, t}^{\mathrm{L}} = 1$. 
$T$ is the given restoration time length, and $\Omega_{L}$ and $\Omega_{\phi}$ are load bus and phase sets separately.
The nodal active power balance of this objective function is:
\begin{equation}
\label{eq: nodal power balance}
\sum_{k \in \Omega_{\mathrm{K}}(i, .)}P_{k, \phi, t}^{\mathrm{K}}-\sum_{k \in \Omega_{\mathrm{K}}(., i)}P_{k, \phi, t}^{\mathrm{K}}=P_{i, \phi, t}^{\mathrm{G}}-x_{i, t}^{\mathrm{L}} P_{i, \phi, t}^{\mathrm{L}}, \forall i, \phi, t
\end{equation}
where $P_{k, \phi, t}^{\mathrm{K}}$ is the active and reactive power flows along line K, and $P_{i, \phi, t}^{\mathrm{G}}$ is the power outputs of the generators. $\Omega_{\mathrm{K}}(i, .)$ and $\Omega_{\mathrm{K}}(., i)$ represent branches flowing out and in bus $i$. 
For some loads with variability, load forecasting results $f_{\theta}(\textbf{X})$ are used as $ P_{i, \phi, t}^{\mathrm{L}}$ to make restoration strategies.
Then these forecasting values are directly engaged in the objective function and power balance constraint.
Furthermore, other constraints such as connectivity constraints and topological constraints can also be indirectly influenced. 
Optimization problem details can be found in \cite{zhang2021two}.


Load behavior after blackouts differs from normal operation due to the CLPU phenomenon.
To capture this, we improve the load model by adopting the exponential decay model:
\begin{equation}
    P_{\mathrm{CLPU}}(t)=P_{0} \cdot(1+a \cdot e^{-\frac{t-t_{0}}{\tau}}), \quad t>0
\end{equation}
where $P_0$ is the power value expected in normal operation. The parameter $\alpha$ is the overshoot value of the CLPU event. The parameter $\tau$ denotes the decay time with $t_0$ being the time of reconnection.

\subsection{Validation Model for the Proposed Adversarial Attack Strategies Using Optimal Power Flow}
\begin{figure}
    \centering
    \includegraphics[width=1\linewidth]{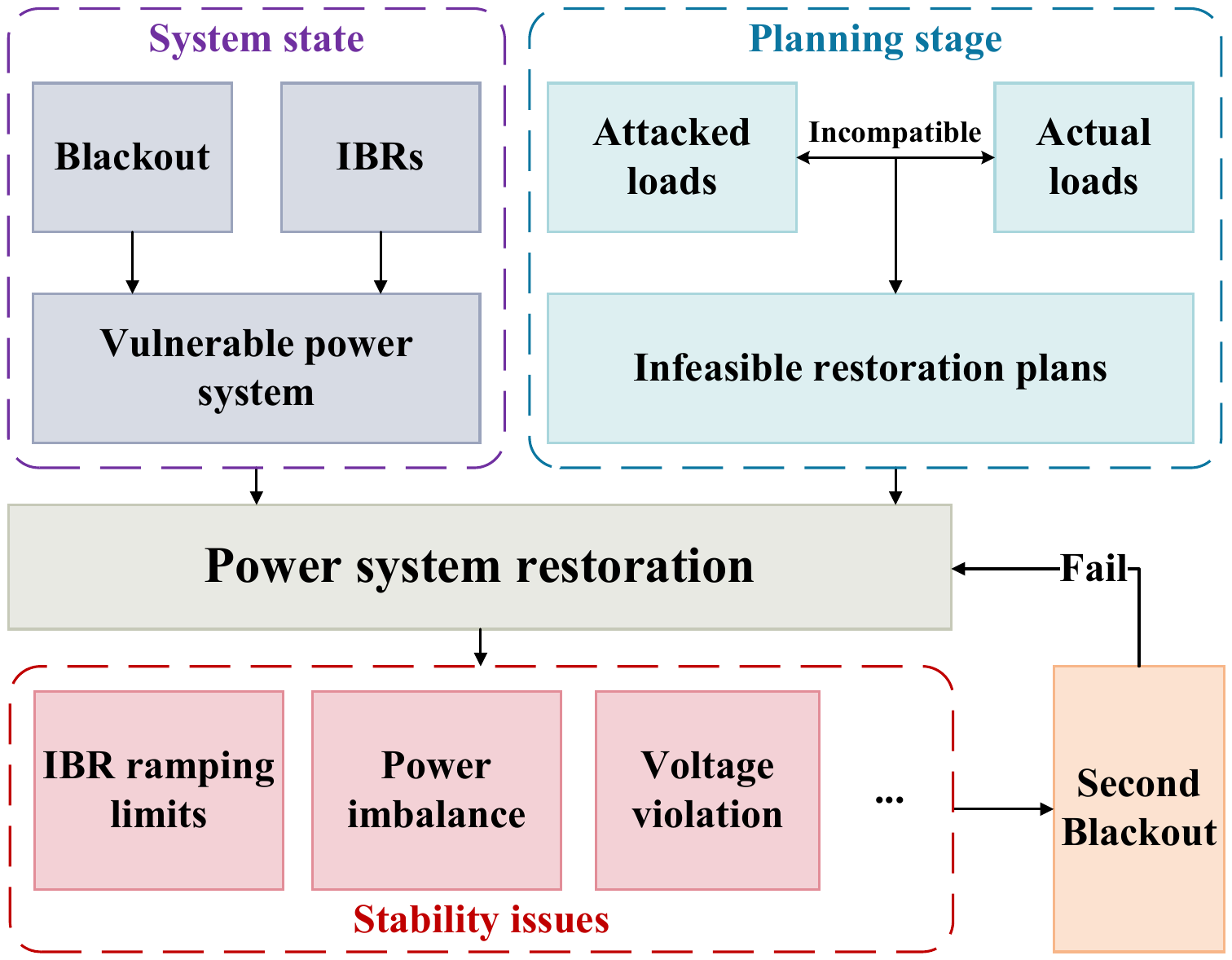}
    \caption{Vulnerabilities of power system restoration under adversarial attacks on load forecasting. System state is vulnerable due to blackout and IBRs, and incompatibility between the attacked load forecasts and actual loads leads to infeasible restoration plans in the planning stage. When applied to the real system, both will cause stability issues, thereby compromising restoration process.}
    \label{fig:restoration vulnerability under attack}
\end{figure}

Fig. \ref{fig:restoration vulnerability under attack} illustrates how adversarial attacks on load forecasting affect real-world restoration. 
First, blackouts and the low-inertia nature of inverter-based resources (IBRs) put the power system in a significantly vulnerable state.
Second, the forecasting loads under adversarial attacks are incompatible with the actual loads, which makes restoration plans based on attacked forecasts infeasible under actual system conditions.
Consequently, active-power-balance infeasibility and ramping-limited IBR response can arise when such plans are applied to a practical system.
This can delay load pickup and make the intended restoration stage unsuccessful.
These safety concerns highlight the vulnerability of restoration planning to adversarial attacks on load forecasting.

To assess the operational feasibility of restoration plans, we propose a three-phase unbalanced OPF validation framework. In this framework, restoration plans are generated based on the attacked forecasts $f_{\theta}(\tilde{\mathbf{X}})$produced by the proposed SAA, but evaluated under the true loads. By analyzing the simulation results, the underlying causes of restoration failures can be identified, which not only quantifies the practical restoration impact of the vulnerable points revealed in SAA but also provides actionable guidance on how system resilience can be enhanced.
Focusing on cyber resilience, this study assumes an accurate load forecasting model.
At each restoration stage $t_s$, the objective function is to maximize the loads with weight factor $w_{i}^{\mathrm{L}}$:
\begin{equation}
    \max \sum_{i \in \Omega_{L}} \sum_{\phi \in \Omega_{\phi}}\left(w_{i}^{\mathrm{L} } P_{i, \phi, t_s}^{\mathrm{L}}\right)
\end{equation}
where  $P_{i, \phi, t_s}^{\mathrm{L}}$ is the restored load value at restoration stage $t_s$.
Constraints (\ref{eq: nodal active power balance}) and (\ref{eq: nodal reactive power balance}) are the nodal power balance constraints, where $P_{k, \phi, t_s}^{\mathrm{K}}$ and $Q_{k, \phi, t_s}^{\mathrm{K}}$ are the active and
reactive power flows along line $k$, and $P_{i, \phi, t_s}^{\mathrm{G}}$ and $Q_{i, \phi, t_s}^{\mathrm{G}}$ are the output power of generators:
\begin{equation}
\label{eq: nodal active power balance}
\begin{array}{c}
    \sum_{k \in \Omega_{\mathrm{K_s}}\left(i, \cdot\right)} P_{k, \phi, t_s}^{\mathrm{K}}-\sum_{k \in \Omega_{\mathrm{K_s}}(\cdot, i)} P_{k, \phi, t_s}^{\mathrm{K}}=P_{i, \phi, t_s}^{\mathrm{G}}-P_{i, \phi, t_s}^{\mathrm{L}}\\
    \forall i \in  \Omega_{\mathrm{B_s}}, \phi 
\end{array}
\end{equation}
\begin{equation}
\label{eq: nodal reactive power balance}
\begin{array}{c}
    \sum_{k \in \Omega_{\mathrm{K_s}}\left(i, \cdot\right)} Q_{k, \phi, t_s}^{\mathrm{K}}-\sum_{k \in \Omega_{\mathrm{K_s}}(\cdot, i)} Q_{k, \phi, t_s}^{\mathrm{K}}=Q_{i, \phi, t_s}^{\mathrm{G}}-Q_{i, \phi, t_s}^{\mathrm{L}}\\
    \forall i \in  \Omega_{\mathrm{B_s}}, \phi 
\end{array}
\end{equation}
$ \Omega_{\mathrm{K_s}}$ and $\Omega_{\mathrm{B_s}}$ are restored lines and buses at restoration stage $t_s$.
Line active and reactive power limits are given in constraints (\ref{eq: line active power capacity}) and (\ref{eq: line reactive power capacity}), where $P_{k}^{\mathrm{K}, \mathrm{M}}$ and $Q_{k}^{\mathrm{K}, \mathrm{M}}$ are maximum allowable active and reactive power:
\begin{equation}
\label{eq: line active power capacity}
    -P_{k}^{\mathrm{K}, \mathrm{M}} \leq P_{k, \phi, t_s}^{\mathrm{K}} \leq P_{k}^{\mathrm{K}, \mathrm{M}}, \forall k \in \Omega_{\mathrm{K_s}}, \phi 
\end{equation}
\begin{equation}
\label{eq: line reactive power capacity}
    -Q_{k}^{\mathrm{K}, \mathrm{M}} \leq Q_{k, \phi, t_s}^{\mathrm{K}} \leq Q_{k}^{\mathrm{K}, \mathrm{M}}, \forall k \in \Omega_{\mathrm{K_s}}, \phi
\end{equation}
Constraint (\ref{eq: voltage limits}) limits voltage $U_{i, \phi, t_s}$ to an allowable range $[U_{i}^{\mathrm{m}}, U_{i}^{\mathrm{M}}]$:
\begin{equation}
\label{eq: voltage limits}
    U_{i}^{\mathrm{m}} \leq U_{i, \phi, t_s} \leq U_{i}^{\mathrm{M}}, \forall i \in \Omega_{\mathrm{B_s}}, \phi
\end{equation}
Constraint (\ref{eq: voltage difference}) calculate the voltage difference of bus $i$ and bus $j$ along the line $k$:
\begin{equation}
\label{eq: voltage difference}
    U_{i, \phi, t_s}-U_{j, \phi, t_s}=2\left(\hat{R}_{k} P_{k, \phi, t_s}^{\mathrm{K}}+\hat{X}_{k} Q_{k, \phi, t_s}^{\mathrm{K}}\right), \forall k, i j \in \Omega_{\mathrm{K_s}}, \phi
\end{equation}
where $\hat{R}_{k}$ and $\hat{X}_{k}$ are the unbalanced three-phase resistance
matrix and reactance matrix of line $k$. More details of these two variables can be found in \cite{zhang2021two}. 
Constraints (\ref{eq: NBS active generation}) and (\ref{eq: NBS reactive generation}) give output power of grid-following (GFL) IBRs:
\begin{equation}
\label{eq: NBS active generation}
    P_{i, \phi, t_s}^{\mathrm{G}} = P_{i, \phi, t_s}^{\mathrm{G}, setpoint}, \forall i \in \Omega_{\mathrm{GFL_s}}, \phi
\end{equation}
\begin{equation}
\label{eq: NBS reactive generation}
    Q_{i, \phi, t_s}^{\mathrm{G}}  = Q_{i, \phi, t_s}^{\mathrm{G}, setpoint}, \forall i \in \Omega_{\mathrm{GFL_s}}, \phi
\end{equation}
where $\Omega_{\mathrm{GFL_s}}$ is the restored GFL IBRs set, and $P_{i, \phi, t_s}^{\mathrm{G}, setpoint}$ and $Q_{i, \phi, t_s}^{\mathrm{G}, setpoint}$ are the calculated power outputs setpoints for these IBRs from results of the optimization problem \eqref{eq: restoration objective}. 
During restoration, the output power of inverter-based GFL IBRs is typically set as a fixed value determined by system operators \cite{gpst2023ibr}.
Based on this, we constrain these IBRs' output to match the dispatch results of the restoration model.
For grid-forming (GFM) IBRs, constraints (\ref{eq: BS initial active power}) and (\ref{eq: BS initial reactive power}) limit the active and reactive power outputs when $t_s=1$:
\begin{equation}
\label{eq: BS initial active power}
    0 \leq P_{i, \phi, t_s}^{\mathrm{G}} \leq P_{i}^{\mathrm{G}, \mathrm{M}}, \forall i \in \Omega_{\mathrm{GFM_s}}, \phi ,t_s=1
\end{equation}
\begin{equation}
\label{eq: BS initial reactive power}
    0 \leq Q_{i, \phi, t_s}^{\mathrm{G}} \leq Q_{i}^{\mathrm{G}, \mathrm{M}}, \forall i \in \Omega_{\mathrm{GFM_s}}, \phi ,t_s=1
\end{equation}
After that, to avoid the potential large frequency deviations caused by MG formation and oversized load restoration, we restrict the generation ramping rate:
\begin{equation}
\label{eq: BS ramping limit}
    \begin{array}{c}
-P_{i, t_s}^{\mathrm{G}, \mathrm{MLS}} \leq P_{i, \phi, t_s}^{\mathrm{G}}-P_{i, \phi, t_s-1}^{\mathrm{G}} \leq P_{i, t_s}^{\mathrm{G}, \mathrm{MLS}}, \\
i \in \Omega_{\mathrm{GFM_s}}, \phi, t_s \geq 2
\end{array}
\end{equation}
where
\begin{equation}
\label{eq: frequency limit}
    \begin{array}{c}
0 \leq P_{i, t_s}^{\mathrm{G}, \mathrm{MLS}} \leq P_{i, t_s-1}^{\mathrm{G}, \mathrm{M
LS}}+\alpha\left(f^{nadir}- f^{min}\right) \\
\forall i \in \Omega_{\mathrm{GFM_s}}, t_s \geq 2
\end{array}
\end{equation}
Here $P_{i, t_s}^{\mathrm{G}, \mathrm{MLS}}$ is the maximum load step. 
The hyper-parameter $\alpha$ serves as a sensitivity coefficient that characterizes the frequency-power response of IBRs. 
$f^{nadir}$ is the lowest frequency when loads are restored to the grid, and $f^{min}$ is a user-defined minimum allowable frequency. In this way, we can ensure the restored load and frequency response do not exceed the user-defined thresholds.
\begin{figure}
    \centering
    \includegraphics[trim=0 10 280 10, clip , width=0.85\linewidth]{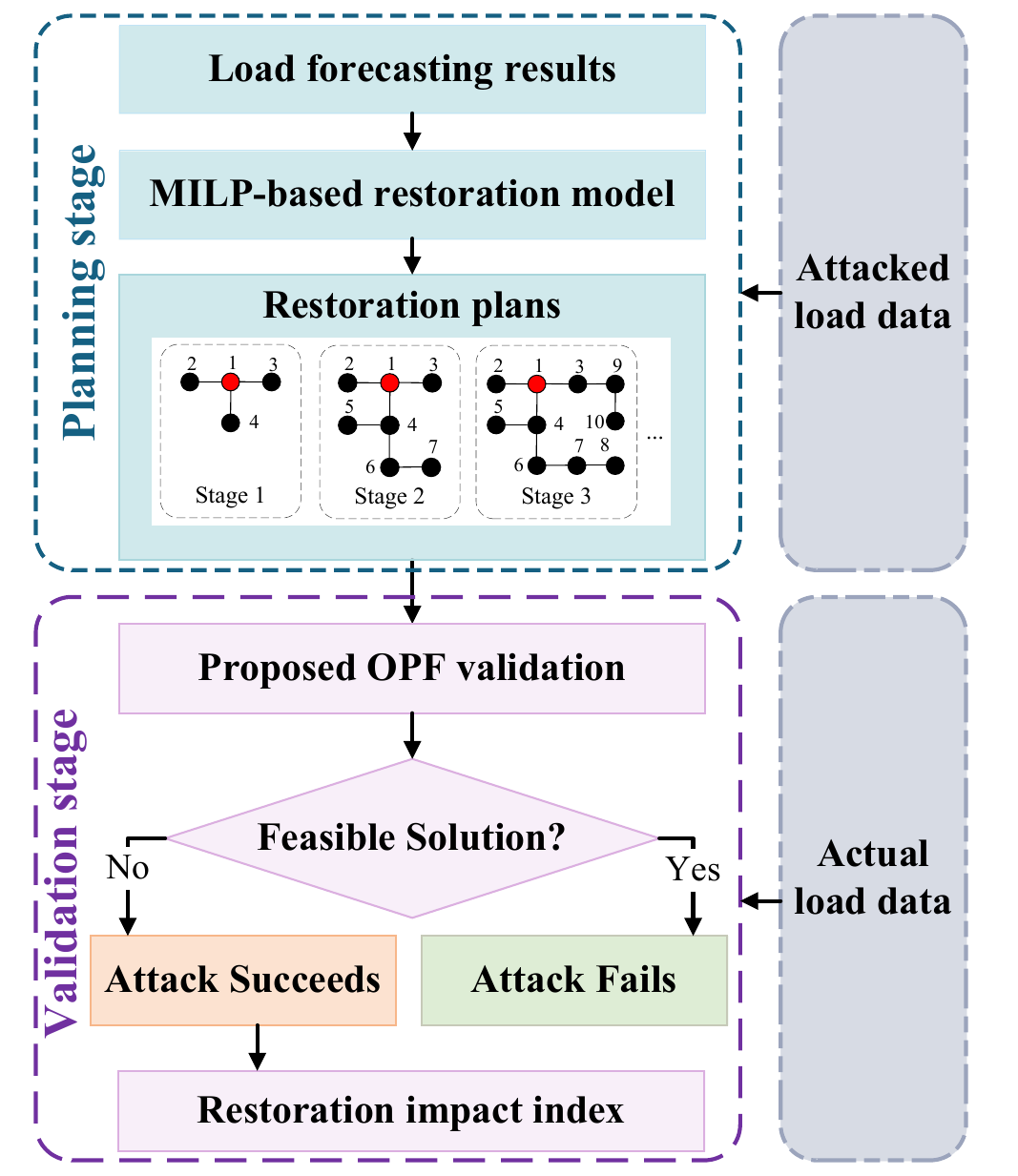}
    \caption{Workflow of cyber resilience evaluation under adversarial attack. In the planning stage, restoration plans are generated from a MILP-based model using attacked load forecasting results. In the validation stage, the proposed OPF checks its feasibility under actual loads. If OPF fails, the results are then used to compute the Restoration impact index.}
    \label{fig: work flow}
\end{figure}

\subsection{Quantifying Cyber Resilience during Restoration}
To quantify how the sparse perturbations generated by SAA affect the resilience of the restoration process, we define a resilience metric named Restoration Impact Index (RII).
Unlike conventional forecasting metrics, the RII is intended to evaluate cyber resilience over the full forecasting-restoration pipeline.
Its key purpose is to measure how severely the restoration process degrades when it is subjected to sparse perturbations.
Specifically, the OPF-based validation framework is first used to determine whether the perturbed forecasts make the restoration plan operationally infeasible, and this process identifies the minimum sparsity required to trigger restoration failure.
The resulting unmet-load ratio is then normalized by that minimum perturbation level, yielding
\begin{equation}
    RII = R_{L}/ k_{min}
\end{equation}
where $k_{min}$ denotes the minimum sparsity level that can cause restoration failure in the OPF validation.
$R_{L}$ represents the ratio of unrestored loads to all demand:
\begin{equation}
    R_{L} = \frac{P_{\text{total}} -  \sum_{i \in \Omega_L} \sum_{\phi \in \Omega_\phi} P_{i,\phi}^L}{P_{\text{total}}}
\end{equation}
$P_{i,\phi}^L$ is the restored load at bus $i$ and phase $\phi$, and $P_{\text{total}}$ is the total demand to be restored.
Thus, $R_{L}$ measures restoration degradation, while $k_{min}$ reflects tolerance to sparse perturbations; their ratio increases when large load loss is caused by fewer perturbed entries.
A larger RII indicates weaker cyber resilience, meaning that the restoration process suffers larger unrestored load and loses feasibility under smaller perturbations.
When $RII=0$, it indicates that all load is restored in the validated restoration outcome.
In this way, the RII provides a quantitative bridge between sparse perturbations and downstream restoration consequences, enabling the comparison among attacks with different sparsity levels.

The workflow of our proposed validation framework is illustrated in Fig. \ref{fig: work flow}.
It contains two stages: planning stage and validation stage.
In the first stage, the MILP-based restoration model generates restoration plans based on the attacked load forecasting results.
In the second stage, these plans are sent to the OPF model to assess their feasibility under actual loads. 
If the OPF has feasible solutions, the attack fails, and the restoration plan remains feasible. Conversely, if no feasible solution is found, the attack is deemed successful, which means the restoration plan is infeasible in practical scenarios.
Then based on the OPF results, the restoration impact index is finally calculated to quantify the restoration consequence of the attack.

\section{Case Study}
\subsection{Experimental Setup Description}
The dataset used for load forecasting contains historical weather data and hourly load profiles for U.S. commercial and residential buildings.
Commercial loads include the most representative commercial building types in the U.S., and residential loads include BASE, HIGH, and LOW residential load variants to approximate different housing conditions. 
The details of this dataset can be found in \cite{OEDI_Dataset_153}.
We use two load forecasting models: LSTM and CNN-LSTM. 
Each model input is a sequence matrix consisting of the 72 hour's historical data.
The attack is implemented in the testing stage to simulate the real world black box attack scenario.

\begin{figure}
    \centering
    \includegraphics[trim=20 80 210 30, clip , width=1\linewidth]{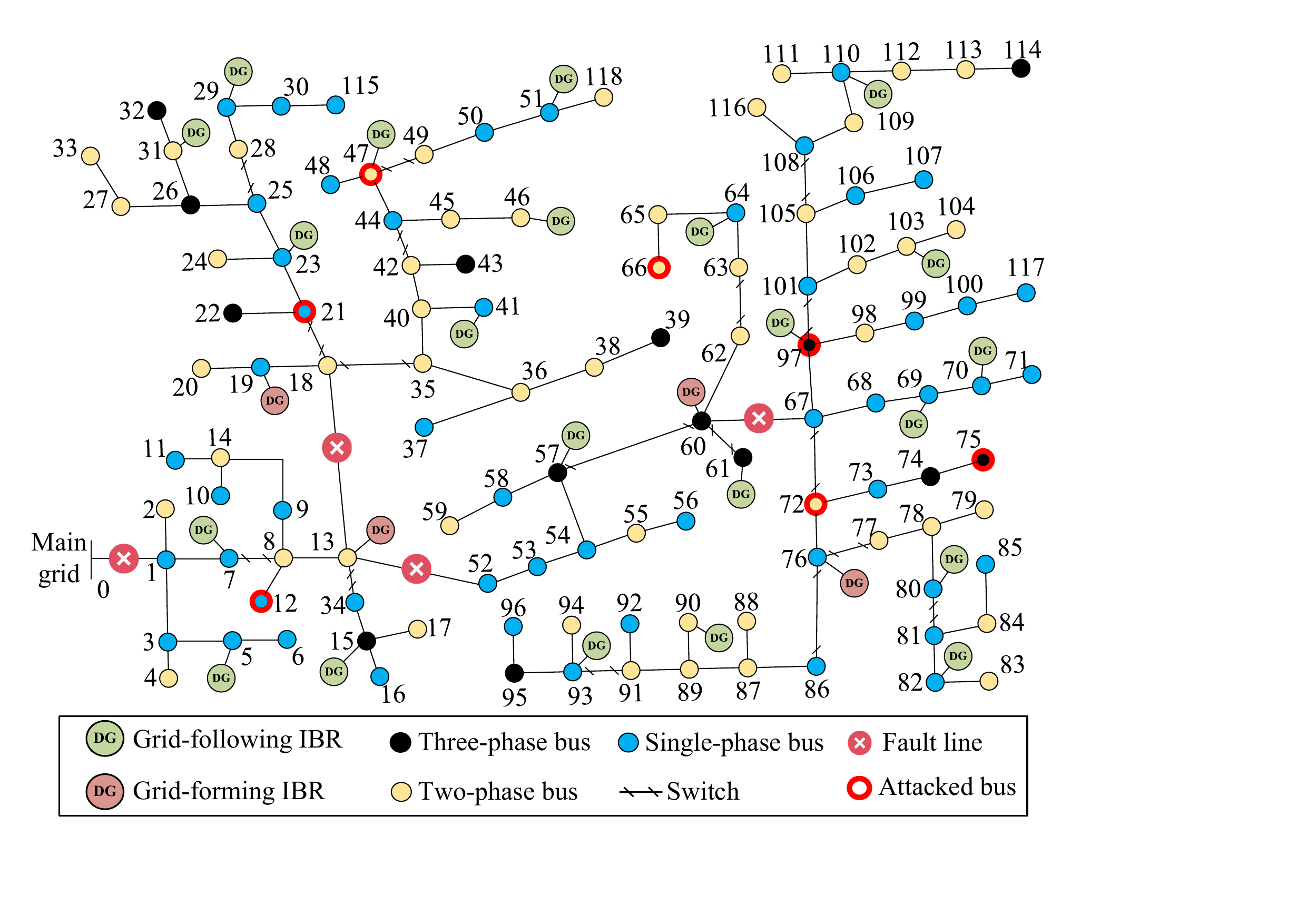}
    \caption{Modified IEEE 123 node test feeder.}
    \label{fig: Modified IEEE 123 node test feeder}
\end{figure}

\begin{table}[h]
\centering
\renewcommand{\arraystretch}{1.5}
\caption{Locations and capacities of GFL and GFM IBRs in IEEE 123 test feeder}
\begin{tabular}{lll}
\Xhline{1.2pt}
\textbf{Type} & \textbf{Locations} & \textbf{Capacities} \\
\hline
\makecell[l]{GFL\\ IBR (1-$\phi$)} & 
\makecell[l]{7, 15, 41, 46, 47, 51,
\\ 61, 64, 69, 82, 93, 97,\\103} & 
\makecell[l]{400 kW for single-phase \\ 200 kVAr for single-phase} \\
\hline
\makecell[l]{GFL\\ IBR (3-$\phi$)} & 
\makecell[l]{5, 23, 29, 31, 57, 70,\\ 80, 90, 110} & 
\makecell[l]{500 kW per phase \\ 250 kVAr per  phase } \\
\hline
\makecell[l]{GFM\\IBR (3-$\phi$)} & 
13, 19, 60, 76 & 
\makecell[l]{500 kW per  phase  \\ 250 kVAr per  phase } \\
\Xhline{1.2pt}
\end{tabular}
\label{tab:IBR_config}
\end{table}

We simulate the restoration process using a modified three-phase unbalanced IEEE 123-bus test system as shown in Fig. \ref{fig: Modified IEEE 123 node test feeder}, where different bus types are color-coded and remote switches are also included.
Table \ref{tab:IBR_config} gives the locations and capacities of GFM and GFL IBRs.
We choose seven buses to implement adversarial attacks during the restoration process, as shown in red circles in Fig. \ref{fig: Modified IEEE 123 node test feeder}.
Instead of normal forecasts $f_{\theta}(\textbf{X})$, the load values in these buses are attacked $f_{\theta}(\tilde{\textbf{X})}$.
With regard to the CLPU model, we present the values of parameters $\alpha$ and $\tau$ for different load types and restoration time in Table \ref{tab:alpha_tau}, which are summarized from  \cite{hachmann2019cold}. 

\begin{table}[h]
\centering
\renewcommand{\arraystretch}{1.2}
\caption{Values of parameters $\alpha$ and $\tau$ in CLPU for different load type and restoration time}
\begin{tabular}{l c c c}
\Xhline{1.2pt}
\textbf{Load Type} & \textbf{Restoration Time} & \textbf{Overshoot} $\boldsymbol{\alpha}$ & \textbf{Time Decay(min)} $\boldsymbol{\tau}$ \\
\hline
\multirow{4}{*}{\makecell[l]{Residential\\Loads}} 
            & Morning   & 1.33 & 11.5 \\
            & Afternoon & 1.03 & 34.0 \\
            & Evening   & 0.62 & 9.8 \\
            & Night     & 0.96 & 10.3 \\
\hline
\multirow{4}{*}{\makecell[l]{Commercial\\Loads}}   & Morning   & 0.62 & 144.1 \\
            & Afternoon & 0.51 & 31.4 \\
            & Evening   & 0.24 & 20.8 \\
            & Night     & 0.70 & 9.4 \\
\Xhline{1.2pt}
\end{tabular}
\label{tab:alpha_tau}
\end{table}


\subsection{Attack Performance on Load Forecasting}

\begin{table*}[t]
\caption{Forecasting Error Increase under PGD and SAA with Different Sparsity Levels}
\label{tab:attack_comparison}
\centering
\renewcommand{\arraystretch}{1.35}
\resizebox{\textwidth}{!}{
\begin{tabular}{llcccccccc}
\Xhline{1.2pt}
\multirow{2}{*}{\textbf{Load Type}} 
& \multirow{2}{*}{\textbf{Dataset}} 
& \multicolumn{4}{c}{\textbf{LSTM}} 
& \multicolumn{4}{c}{\textbf{CNN-LSTM}} \\
\cmidrule(lr){3-6}
\cmidrule(lr){7-10}
& 
& PGD 
& SAA(n=10) 
& SAA(n=50) 
& SAA(n=100)
& PGD 
& SAA(n=10) 
& SAA(n=50) 
& SAA(n=100) \\
\Xhline{0.1pt}

\multirow{4}{*}{Commercial Loads} 
& FullSrvcRestaurant 
& 0.7757 & 2.2637 & 3.6933 & \textbf{3.9839}
& 1.1833 & 3.9088 & 6.0275 & \textbf{6.2012} \\

& MidriseApartment 
& 0.5690 & 1.1693 & 2.7357 & \textbf{3.4420}
& 0.3984 & 1.1654 & 1.8988 & \textbf{2.0962} \\

& QuickSrvcRestaurant 
& 0.3273 & 1.2073 & 1.9398 & \textbf{2.0225}
& 0.6021 & 2.2967 & 3.2306 & \textbf{3.2915} \\

& SmallHotel 
& 0.7192 & 1.4080 & 3.1012 & \textbf{3.8237}
& 0.8487 & 3.6117 & 5.7933 & \textbf{6.2246} \\

\Xhline{0.1pt}

\multirow{3}{*}{Residential Loads} 
& BASE 
& 0.1149 & 0.2628 & 0.5978 & \textbf{0.7440}
& 0.1699 & 0.4193 & 0.7994 & \textbf{0.9185} \\

& HIGH 
& 0.2333 & 0.3459 & 0.7836 & \textbf{0.9769}
& 0.2748 & 0.6079 & 1.0783 & \textbf{1.2689} \\

& LOW 
& 0.0603 & 0.1038 & 0.2427 & \textbf{0.3005}
& 0.0709 & 0.1660 & 0.3473 & \textbf{0.4001} \\

\Xhline{1.2pt}
\end{tabular}
}
\end{table*}

Table \ref{tab:attack_comparison} compares the proposed SAA with PGD across LSTM and CNN-LSTM forecasting models. The values represent the MSE increase relative to the clean condition under the same perturbation bound and iteration setting. PGD is used as a standard gradient-based baseline, whereas SAA perturbs only the top-$n$ spatiotemporal weather-input entries selected by gradient magnitude.
The results show that sparsity does not weaken the attack; instead, targeted sparsity improves its effectiveness. Even with $n=10$, SAA produces larger error increases than PGD for every tested dataset. On average, SAA($n=10$) is about 2.41 times stronger than PGD for LSTM and 3.43 times stronger for CNN-LSTM. This suggests that forecasting vulnerability is concentrated in a small subset of weather-input coordinates rather than being uniformly distributed across the input matrix.
As $n$ increases from 10 to 100, the MSE increase continues to grow, but the marginal gain becomes smaller. This diminishing trend indicates that the influential coordinates are limited, and additional perturbations mainly refine rather than fundamentally change the attack effect. Commercial loads exhibit larger degradation than residential loads, suggesting stronger weather dependence or sharper temporal patterns. CNN-LSTM is also more sensitive than LSTM in most cases, which may be caused by convolutional feature extraction amplifying localized perturbations. These results support using SAA as a vulnerability-screening tool before forecast outputs are used in restoration planning.

\subsection{RII Evaluation on Microgrids}


\begin{table}[t]
\centering
\caption{RII results under SAA and PGD attacks for different MGs}
\label{tab:rii_results_saa_pgd}
\renewcommand{\arraystretch}{1.2}
\setlength{\tabcolsep}{3.5pt}
\footnotesize
\begin{tabular}{c c c c c c c}
\toprule
MG & Attacked Bus & Attack & Failed Stage & Sparsity & $R_L$ & RII \\
\midrule
\multirow{2}{*}{1} 
& \multirow{2}{*}{12} 
& SAA & -- & -- & 0 & 0 \\
& 
& PGD & -- & 72 & 0 & 0 \\
\midrule
\multirow{2}{*}{2} 
& \multirow{2}{*}{21, 47} 
& SAA & 3 & 6 & 0.215979 & 0.035996 \\
& 
& PGD & 3 & 72 & 0.213403 & 0.002963 \\
\midrule
\multirow{2}{*}{3} 
& \multirow{2}{*}{66} 
& SAA & 3 & 8 & 0.012130 & 0.001516 \\
& 
& PGD & -- & 72 & 0 & 0 \\
\midrule
\multirow{2}{*}{4} 
& \multirow{2}{*}{72, 75, 97} 
& SAA & 2 & 4 & 0.810421 & 0.202605 \\
& 
& PGD & 2 & 72 & 0.863140 & 0.011988 \\
\bottomrule
\end{tabular}
\end{table}

Table \ref{tab:rii_results_saa_pgd} compares restoration impacts under SAA and PGD. The sparsity column reports the minimum number of perturbed entries $k_{\min}$ required to trigger failure for SAA and the fixed perturbation budget $k=72$ for PGD. Therefore, the comparison reflects not only failure severity, measured by $R_L$, but also attack efficiency, measured by how many input entries are needed to induce infeasibility.
SAA produces comparable or stronger restoration impacts with far fewer perturbations. In MG 2, both attacks cause Stage-3 failure with similar unrestored-load ratios, but SAA needs only 6 entries instead of 72, yielding a much higher RII. In MG 3, SAA triggers Stage-3 failure while PGD remains feasible, showing that vulnerability depends more on the perturbed coordinates than on the total budget. MG 4 is the most critical case: both attacks cause Stage-2 failure, but SAA reaches infeasibility with only four entries and the highest RII (0.202605). MG 1 remains feasible under both attacks, suggesting sufficient GFM balancing margin or limited downstream influence from the attacked bus.


\subsection{Power System Restoration and OPF Validation Results Under Adversarial Attack}
\begin{figure}
    \centering
    \includegraphics[trim=20 70 170 30, clip , width=1\linewidth]{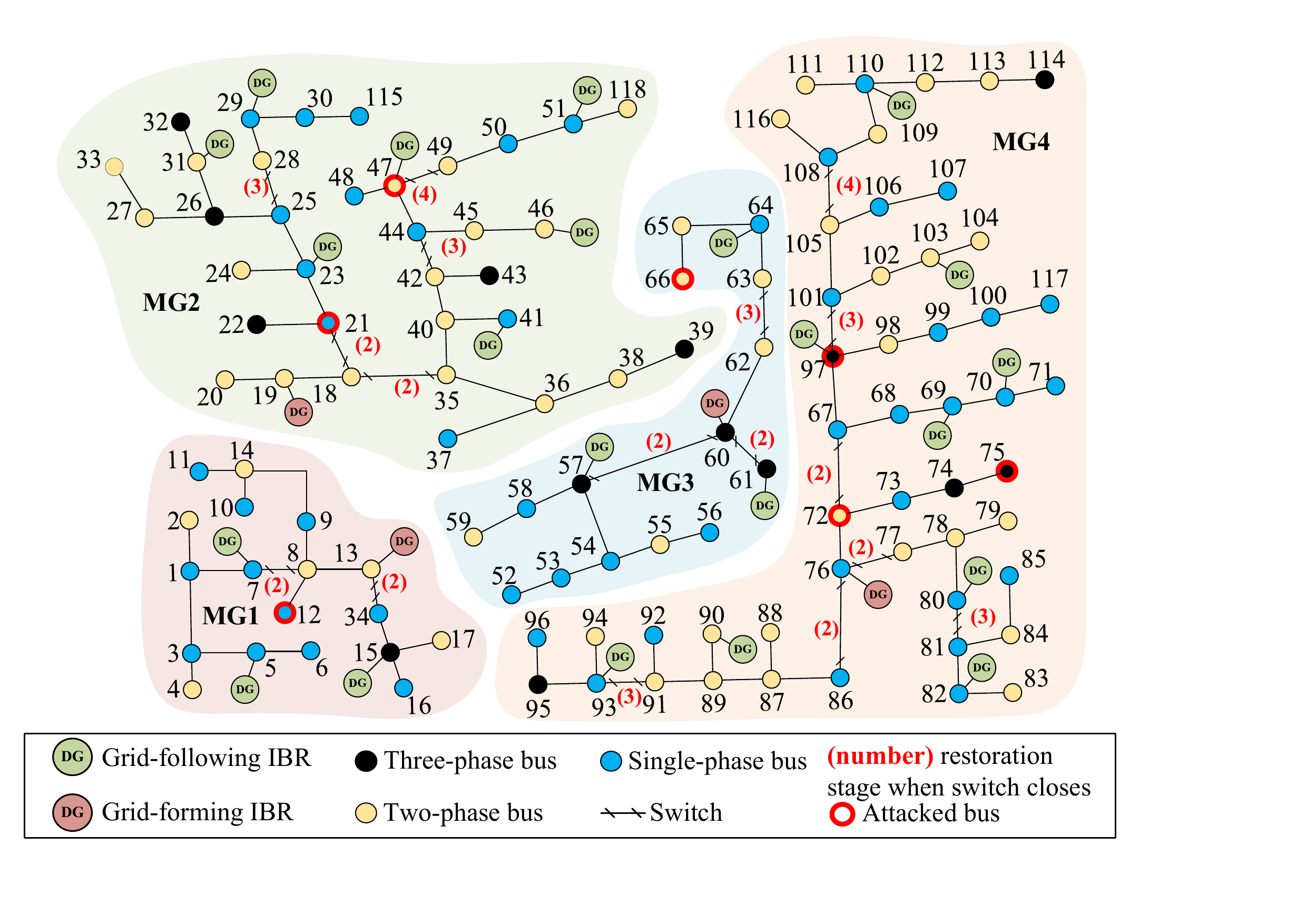}
    \caption{Restoration results of the IEEE 123 bus test feeder, where the restoration stage when the line switch closes is shown in red.}
    \label{fig: restoration sequence}
\end{figure}
\begin{figure}
    \centering
    \includegraphics[trim=10 10 2250 35, clip ,width=0.95\linewidth]{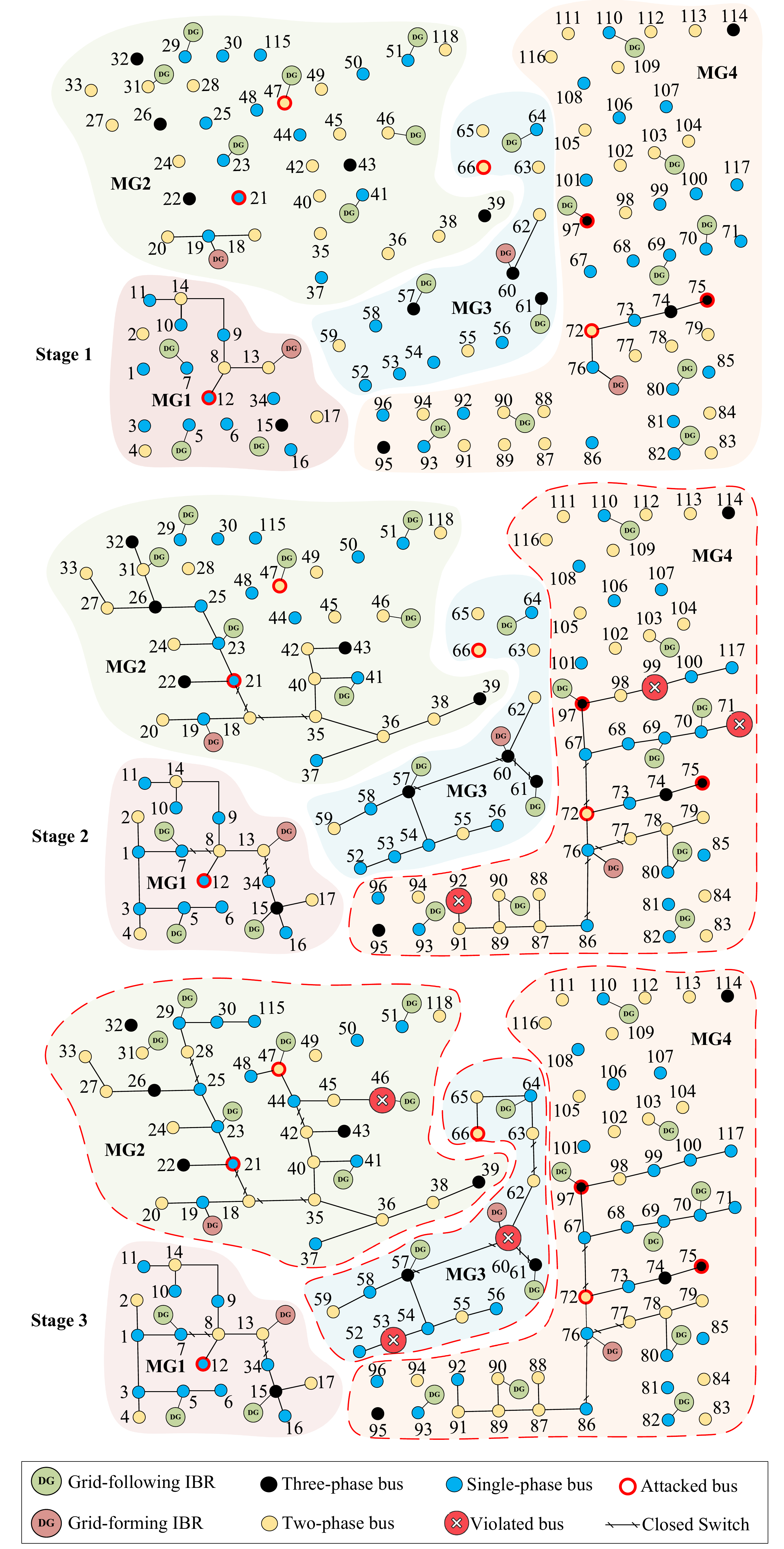}
    \caption{OPF Validation of the proposed SAA. The microgrid in red dashed lines fails the OPF validation in the corresponding stages. MG 1 restores successfully despite attacks, while MG 4 fails in stage 2 and MG 2 and 3 fail in stage 3.}
    \label{fig:OPF Validation}
\end{figure}

This subsection illustrates the restoration sequence and OPF validation results under sparsity level $k=8$.
As is shown in Fig. \ref{fig: restoration sequence}, the system is sectionalized into four MGs according to the fault lines and then each MG is recovered vis sequential closure of switchable lines.
The red numbers in red brackets adjacent to each line switch indicate the corresponding restoration stage at which the switch is closed.
Table \ref{tab:restored_loads} gives the restored sequences of the IBRs and loads, where the subscript is the bus index. '\--' in the table represents that no generators or loads are restored in the current stage.
While MG 1 and MG 3 take two and three stages to be fully restored, respectively, MG 2 and MG 4 need four stages.
We also schedule the restoration sequence based on the normal forecasting loads $f_\theta(\textbf{X})$, and the resulting switching and load-pickup order is the same as in Table \ref{tab:restored_loads}. This indicates that the sparse attack does not necessarily alter the topological restoration sequence; instead, its main impact appears in the OPF feasibility when the attacked plan is tested under actual loads.
\begin{table}
\centering
\caption{Restoration Sequences of IBRs and Loads in Each Microgrid Across 4 Stages Under Adversarial Attack}
\label{tab:restored_loads}
\renewcommand{\arraystretch}{1.4}
\begin{tabular}{ccll}
\toprule
\textbf{MG} & \textbf{Stage} & \textbf{Restored IBRs} & \textbf{Restored Loads} \\
\midrule
\multirow{4}{*}{1} 
& 1 & $G_{13}$ & \makecell[l]{$L_{8}, L_{9}, L_{10}, L_{11}, L_{12}, L_{13}, $ \\ $L_{14}$ }\\
\cline{2-4}
 & 2 & $G_{5}, G_{7}$ &\rule{0pt}{13pt}\makecell[l]{$L_{1}, L_{2}, L_{3}, L_{4}, L_{5}, L_{6},$\\$ L_{7}, L_{15}, L_{16}, L_{17}$} \\
\cline{2-4}
& 3 & \textit{Restoration Complete }& \\
\cline{2-4}
& 4 & \textit{Restoration Complete } &  \\
\midrule
\multirow{4}{*}{2} 
& 1 & $G_{19}$ & $L_{18}, L_{19}, L_{20}$ \\
\cline{2-4}
& 2 & $G_{23}, G_{31}, G_{41}$ &\rule{0pt}{22pt}\makecell[l]{$L_{21}, L_{22}, L_{23}, L_{24}, L_{25}, L_{26},$\\$ L_{27}, L_{31}, L_{32}, L_{33}, L_{35}, L_{36},$\\$ L_{37}, L_{38}, L_{39}, L_{40}, L_{41}, L_{42},$\\$ L_{43}$}  \\
\cline{2-4}
& 3 & $G_{46}$ &\rule{0pt}{13pt}\makecell[l]{$L_{28}, L_{29}, L_{30}, L_{44}, L_{45}, L_{46},$\\$ L_{47}, L_{48}, L_{115}$} \\
\cline{2-4}
& 4 & $G_{51}$ & $L_{49}, L_{50}, L_{51}, L_{118}$ \\
\midrule
\multirow{4}{*}{3} 
& 1 & $G_{60}$ & $L_{60}, L_{62}$ \\
\cline{2-4}
& 2 & $G_{57}, G_{61}$ &\rule{0pt}{13pt}\makecell[l]{$L_{52}, L_{53}, L_{54}, L_{55}, L_{56}, L_{57},$\\$ L_{58}, L_{59}, L_{61}$} \\
\cline{2-4}
& 3 & -- & $L_{63}, L_{64}, L_{65}, L_{66}$ \\
\cline{2-4}
& 4 & \textit{Restoration Complete } & \\
\midrule
\multirow{4}{*}{4} 
& 1 & $G_{76}$ & $L_{72}, L_{73}, L_{74}, L_{75}, L_{76}$ \\
\cline{2-4}
& 2 & $G_{69}, G_{70}, G_{90}, G_{97}$ &\rule{0pt}{22pt}\makecell[l]{$L_{67}, L_{68}, L_{69}, L_{70}, L_{71}, L_{77},$\\$ L_{78}, L_{79}, L_{80}, L_{86}, L_{87}, L_{88},$\\$ L_{89}, L_{90}, L_{91}, L_{92}, L_{97}, L_{98},$\\$ L_{99}, L_{100}, L_{117}$ }\\
\cline{2-4}
& 3 & $G_{80}, G_{82}$ &\rule{0pt}{17pt}\makecell[l]{$L_{81}, L_{82}, L_{83}, L_{84}, L_{85}, L_{93},$\\$ L_{94}, L_{95}, L_{96}, L_{101}, L_{102}, $\\$ L_{103}, L_{104}, L_{105}, L_{106}, L_{107}$} \\
\cline{2-4}
& 4 & $G_{93}, G_{110}$ &\rule{0pt}{13pt}\makecell[l]{$L_{108}, L_{109}, L_{110}, L_{111}, L_{112}, $\\$ L_{113}, L_{114}, L_{116}$} \\
\bottomrule
\end{tabular}
\end{table}
Fig. \ref{fig:OPF Validation} depicts the results across three restoration stages; MGs enclosed by red dashed lines fail OPF validation in the corresponding stages.
All microgrids restore successfully in the initial stage.
In stage 2, buses 71, 92, and 99 in MG 4 are reported to violate certain constraints in the OPF.
Although violations do not happen for MG 3 in the first two stages, buses 60 and 53 meet problems when the attacked bus 66 is restored at stage 3. 
With regard to MG 2, while load change caused by the attacked bus 21 can be balanced by IBRs, restored bus 46 still influences the restoration process in stage 3.

Table \ref{tab:violated_constraints} summarizes the constraints that block feasibility and would require relaxation to recover feasible OPF solutions.
As shown, the primary cause is the violation of the nodal active power balance constraint (\ref{eq: nodal power balance}).
In concrete, MG 4 fails at Stage 2 due to power balance violations at buses 75 $\phi_c$, 92 $\phi_a$, and 99 $\phi_a$. Similarly, MG 2 fails at Stage 3 with an imbalance at bus 46 $\phi_a$, while MG 3 experiences failures at buses 23 $\phi_a$ and 60 $\phi_a$ in the same stage. 
These results illustrate that even a few buses under attack can propagate constraint violations throughout the network, rendering the restoration plan infeasible.

\begin{table}
\centering
\caption{Violated Constraints During Restoration Stages}
\label{tab:violated_constraints}
\renewcommand{\arraystretch}{1.4}
\begin{tabular}{ccp{4cm}}
\hline
\textbf{Restoration Stage} & \textbf{Microgrid} & \textbf{Violated Constraints} \\ 
\hline
Stage 2 & MG 4 &Active Power Balance in bus 75 $\phi_c$,  bus 92 $\phi_a$, and bus 99 $\phi_a$ \\ 
\hline
\multirow{2}{*}{Stage 3} & MG 2 &Active Power Balance in bus 46 $\phi_a$\\ 
\cline{2-3}
 & MG 3 &Active Power Balance in bus 23 $\phi_a$ and bus 60 $\phi_a$\\ 
\hline
\end{tabular}
\end{table}

\subsection{Resilience-enhancement Strategies for Failed Restoration Stages}

Although bus 12 is attacked, MG 1 is still energized in the initial stage. 
In this phase, constraint (\ref{eq: BS initial active power}) only limits the upper power output bound for GFM IBRs, allowing sufficient flexibility to handle load deviations.
MG 4 follows a similar pattern, where two attacked buses are restored at stage 1. 
Table \ref{tab:gfm_generation_load} presents the GFM generation and load values.
As shown, even if the load deviation for MG 4 is 21 kW, it can still be balanced. 
To maintain such flexibility in practice, different generation weights can be assigned to IBRs across restoration stages. By assigning higher weights to GFLs in early stages, they can contribute more power so that GFMs retain great adaptability for handling uncertainties in later stages.

On the contrary, MG 4 restoration fails at Stage 2 when bus 97 is energized.
At this time, constraints (\ref{eq: BS ramping limit}) and (\ref{eq: frequency limit}) limit the ramping speed of the GFM IBRs.
Table \ref{tab:gfm_generation_load} shows that the load deviation from 2822 kW to 2844 kW exceeds GFM capabilities.
Similarly, for MG 2 and 3 at stage 3, available GFM generation remains insufficient to handle the load changes in actual conditions.
As load demand surpasses generation capacity, restoration failures occur.
These results show that restoration failures depend on both load-deviation scale and available balancing capacity.
The restoration timing of attacked buses can also critically affect outcomes, suggesting that adversarial perturbations should be incorporated into chance-constrained restoration planning to address cyber-induced uncertainty.
In conclusion, our results highlight the vulnerability of restoration planning to forecasting attacks, especially in low-inertia systems with IBRs.
Maintaining dynamic reserves and applying robust planning strategies are necessary to ensure secure and resilient restoration under cyber attacks.

\begin{table}
\centering
\caption{Comparison of Planned and Actual GFM IBR Generation and Attacked and Actual Loads}
\label{tab:gfm_generation_load}
\begin{tabular}{cccccc}
\toprule
 & \textbf{Stage} & \makecell[c]{\textbf{Planned}\\\textbf{GFM IBR}\\\textbf{Generation}} & \makecell[c]{\textbf{Actual}\\\textbf{GFM IBR}\\\textbf{Generation}} & \makecell[c]{\textbf{Attacked}\\\textbf{Load}\\\textbf{Value}}& \makecell[c]{\textbf{Actual}\\\textbf{Load}\\ \textbf{Value}} \\
\midrule
\multirow{3}{*}{MG 1} & 1 & 631.0 & 634.0 & 631.0 & 634.0 \\
                    & 2 & 641.5 & 642.5 & 1371.0 & 1372.0 \\
                    & 3 & 610.0 & 611.0 & 1037.0 & 1038.0 \\
\midrule
\multirow{3}{*}{MG 2} & 1 & 371.0 & 371.0 & 371.0 & 371.0 \\
                    & 2 & 402.5 & 399.5 & 2421.0 & 2418.0 \\
                    & \textbf{3} & 413.0 & 413.0 & 3004.0 & 3013.0 \\
\midrule
\multirow{3}{*}{MG 3} & 1 & 292.0 & 292.0 & 292.0 & 292.0 \\
                    & 2 & 260.5 & 260.5 & 1303.0 & 1303.0 \\
                    & \textbf{3} & 271.0 & 273.0 & 1469.0 & 1475.0 \\
\midrule
\multirow{3}{*}{MG 4} & 1 & 648.0 & 669.0 & 648.0 & 669.0 \\
                    & \textbf{2} & 637.5 & 658.5 & 2822.0 & 2844.0 \\
                    & 3 & - & - & - & - \\
\bottomrule
\end{tabular}
\end{table}

\section{Conclusion and Future Work}
This paper presents a cyber-resilience analysis of distribution system restoration under sparse adversarial attacks on load forecasting. 
We develop a gradient-based sparse attack strategy that perturbs only the most sensitive input elements , and then embed the attacked forecasts into a sequential MILP-based restoration planner. 
To evaluate the physical consequence of these attacked plans, we further validate the resulting restoration schedules using a three-phase unbalanced OPF model under actual load conditions.

Simulation results show that the sparse perturbations can induce forecasting errors and trigger restoration failures in specific stages. 
These failures are primarily due to violations of nodal power balance caused by strictly-restricted generator ramping capacity.
Our findings underscore the need to incorporate cybersecurity considerations into future restoration planning frameworks, especially as AI-driven load forecasting becomes standard.

\bibliographystyle{IEEETrans}
\bibliography{sample}

@article{ibrahim2022cloud,
  title={Cloud-based smart grids: opportunities and challenges},
  author={Ibrahim, Nehad M and Musleh, Dhiaa and Khan, Mohammed Aftab A and Chabani, Sghaier and Dash, Sujata},
  journal={Biologically inspired techniques in many criteria decision making: proceedings of BITMDM 2021},
  pages={1--13},
  year={2022},
  publisher={Springer}
}

@techreport{nist_ot_2023,
  author = {Stouffer, Keith and others},
  title = {Guide to Operational Technology (OT) Security},
  institution = {National Institute of Standards and Technology},
  number = {NIST SP 800-82 Rev. 3},
  year = {2023},
  month = sep,
  doi = {10.6028/NIST.SP.800-82r3}
}

@techreport{epri_ps39a_2025,
  author       = {{Electric Power Research Institute}},
  title        = {2025 Research Portfolio, Project Set PS39A: Real-time Operations and Situational Awareness},
  institution  = {Electric Power Research Institute},
  address      = {Palo Alto, CA, USA},
  year         = {2025},
  url          = {https://restservice.epri.com/publicattachment/91911},
}

@techreport{nist_api_protection,
  author      = {Ramaswamy Chandramouli and Zack Butcher},
  title       = {Guidelines for API Protection for Cloud-Native Systems},
  institution = {National Institute of Standards and Technology},
  number      = {NIST SP 800-228},
  year        = {2025},
  doi         = {10.6028/NIST.SP.800-228}
}

@misc{owasp_tls,
  author       = {{OWASP Foundation}},
  title        = {Testing for Weak Transport Layer Security},
  year         = {2025},
  note         = {OWASP Web Security Testing Guide},

}

@article{maharjan2025distribution,
  title={Distribution System Blackstart and Restoration Using DERs and Dynamically Formed Microgrids},
  author={Maharjan, Salish and Bai, Cong and Wang, Han and Yao, Yiyun and Ding, Fei and Wang, Zhaoyu},
  journal={IEEE Transactions on Smart Grid},
  year={2025},
  publisher={IEEE}
}

@article{fazlhashemi2025distribution,
  title={Distribution Service Restoration: A Distributionally Robust Reinforcement Learning Approach},
  author={Fazlhashemi, Seyed Saeed and Khodayar, Mohammad E and Khodayar, Mahdi and Wang, Jianhui},
  journal={IEEE Transactions on Smart Grid},
  year={2025},
  publisher={IEEE}
}

@article{vu2023multi,
  title={Multi-agent deep reinforcement learning for distributed load restoration},
  author={Vu, Linh and Vu, Tuyen and Vu, Thanh Long and Srivastava, Anurag},
  journal={IEEE Transactions on Smart Grid},
  volume={15},
  number={2},
  pages={1749--1760},
  year={2023},
  publisher={IEEE}
}

@article{zhang2024vulnerability,
  title={Vulnerability of machine learning approaches applied in iot-based smart grid: A review},
  author={Zhang, Zhenyong and Liu, Mengxiang and Sun, Mingyang and Deng, Ruilong and Cheng, Peng and Niyato, Dusit and Chow, Mo-Yuen and Chen, Jiming},
  journal={IEEE Internet of Things Journal},
  volume={11},
  number={11},
  pages={18951--18975},
  year={2024},
  publisher={IEEE}
}

@article{edib2023situation,
  title={Situation-aware load restoration considering uncertainty and correlation},
  author={Edib, Shamsun Nahar and Lin, Yuzhang and Vokkarane, Vinod M and Qiu, Feng and Zhang, Yichen and Du, Pengwei},
  journal={IEEE Transactions on Power Systems},
  volume={39},
  number={2},
  pages={2611--2629},
  year={2023},
  publisher={IEEE}
}

@article{konar2023mpc,
  title={Mpc-based black start and restoration for resilient DER-rich electric distribution system},
  author={Konar, Srayashi and Srivastava, Anurag K},
  journal={IEEE Access},
  volume={11},
  pages={69177--69189},
  year={2023},
  publisher={IEEE}
}

@article{song2020robust,
  title={Robust distribution system load restoration with time-dependent cold load pickup},
  author={Song, Meng and Sun, Wei and others},
  journal={IEEE Transactions on Power Systems},
  volume={36},
  number={4},
  pages={3204--3215},
  year={2020},
  publisher={IEEE}
}

@article{xie2023dynamic,
  title={Dynamic frequency-constrained load restoration considering multi-phase cold load pickup behaviors},
  author={Xie, Dunjian and Xu, Yan and Nadarajan, Sivakumar and Viswanathan, Vaiyapuri and Gupta, Amit Kumar},
  journal={IEEE Transactions on Power Systems},
  volume={39},
  number={1},
  pages={107--118},
  year={2023},
  publisher={IEEE}
}

@inproceedings{gu2013bad,
  title={Bad data detection method for smart grids based on distributed state estimation},
  author={Gu, Yun and Liu, Ting and Wang, Dai and Guan, Xiaohong and Xu, Zhanbo},
  booktitle={2013 IEEE International Conference on Communications (ICC)},
  pages={4483--4487},
  year={2013},
  organization={IEEE}
}

@techreport{gpst2023ibr,
author  = {Aurecon},
  title        = {The Role of Inverter-Based Resources During System Restoration},
institution = {CSIRO},
  year         = {2023},
  month        = {Jun},
  type         = {Final Report},
    note         = {Online},
  url          = {https://www.csiro.au/-/media/EF/Files/GPST-Roadmap/Stage3-Final/Topic-5_The-role-of-inverter-based-resources.pdf}
}

@incollection{kurakin2018adversarial,
  title={Adversarial examples in the physical world},
  author={Kurakin, Alexey and Goodfellow, Ian J and Bengio, Samy},
  booktitle={Artificial intelligence safety and security},
  pages={99--112},
  year={2018},
  publisher={Chapman and Hall/CRC}
}

@article{chakraborty2021survey,
  title={A survey on adversarial attacks and defences},
  author={Chakraborty, Anirban and Alam, Manaar and Dey, Vishal and Chattopadhyay, Anupam and Mukhopadhyay, Debdeep},
  journal={CAAI Transactions on Intelligence Technology},
  volume={6},
  number={1},
  pages={25--45},
  year={2021},
  publisher={Wiley Online Library}
}

@article{liu2024enhancing,
  title={Enhancing cyber-resiliency of der-based smart grid: A survey},
  author={Liu, Mengxiang and Teng, Fei and Zhang, Zhenyong and Ge, Pudong and Sun, Mingyang and Deng, Ruilong and Cheng, Peng and Chen, Jiming},
  journal={IEEE Transactions on Smart Grid},
  year={2024},
  publisher={IEEE}
}

@article{liu2016power,
  title={Power system restoration: a literature review from 2006 to 2016},
  author={Liu, Yutian and Fan, Rui and Terzija, Vladimir},
  journal={Journal of Modern Power Systems and Clean Energy},
  volume={4},
  number={3},
  pages={332--341},
  year={2016},
  publisher={SGEPRI}
}

@article{chen2017modernizing,
  title={Modernizing distribution system restoration to achieve grid resiliency against extreme weather events: An integrated solution},
  author={Chen, Chen and Wang, Jianhui and Ton, Dan},
  journal={Proceedings of the IEEE},
  volume={105},
  number={7},
  pages={1267--1288},
  year={2017},
  publisher={IEEE}
}

@article{cui2019machine,
  title={Machine learning-based anomaly detection for load forecasting under cyberattacks},
  author={Cui, Mingjian and Wang, Jianhui and Yue, Meng},
  journal={IEEE Transactions on Smart Grid},
  volume={10},
  number={5},
  pages={5724--5734},
  year={2019},
  publisher={IEEE}
}

@inproceedings{chen2018machine,
  title={Is machine learning in power systems vulnerable?},
  author={Chen, Yize and Tan, Yushi and Deka, Deepjyoti},
  booktitle={2018 IEEE International Conference on Communications, Control, and Computing Technologies for Smart Grids (SmartGridComm)},
  pages={1--6},
  year={2018},
  organization={IEEE}
}

@article{zhou2022robust,
  title={Robust load forecasting towards adversarial attacks via Bayesian learning},
  author={Zhou, Yihong and Ding, Zhaohao and Wen, Qingsong and Wang, Yi},
  journal={IEEE Transactions on Power Systems},
  volume={38},
  number={2},
  pages={1445--1459},
  year={2022},
  publisher={IEEE}
}

@inproceedings{chen2019exploiting,
  title={Exploiting vulnerabilities of load forecasting through adversarial attacks},
  author={Chen, Yize and Tan, Yushi and Zhang, Baosen},
  booktitle={Proceedings of the tenth ACM international conference on future energy systems},
  pages={1--11},
  year={2019}
}

@article{fan2024optimizing,
  title={Optimizing attention in a Transformer for multihorizon, multienergy load forecasting in integrated energy systems},
  author={Fan, Jili and Zhuang, Wei and Xia, Min and Fang, WenXuan and Liu, Jun},
  journal={IEEE Transactions on Industrial Informatics},
  year={2024},
  publisher={IEEE}
}

@article{kong2017short,
  title={Short-term residential load forecasting based on LSTM recurrent neural network},
  author={Kong, Weicong and Dong, Zhao Yang and Jia, Youwei and Hill, David J and Xu, Yan and Zhang, Yuan},
  journal={IEEE transactions on smart grid},
  volume={10},
  number={1},
  pages={841--851},
  year={2017},
  publisher={IEEE}
}

@article{tian2018deep,
  title={A deep neural network model for short-term load forecast based on long short-term memory network and convolutional neural network},
  author={Tian, Chujie and Ma, Jian and Zhang, Chunhong and Zhan, Panpan},
  journal={Energies},
  volume={11},
  number={12},
  pages={3493},
  year={2018},
  publisher={MDPI}
}

@article{hachmann2019cold,
  title={Cold load pickup model parameters based on measurements in distribution systems},
  author={Hachmann, Christian and Lammert, Gustav and Hamann, Lucas and Braun, Martin},
  journal={IET Generation, Transmission \& Distribution},
  volume={13},
  number={23},
  pages={5387--5395},
  year={2019},
  publisher={Wiley Online Library}
}

@article{zhang2021two,
  title={A two-level simulation-assisted sequential distribution system restoration model with frequency dynamics constraints},
  author={Zhang, Qianzhi and Ma, Zixiao and Zhu, Yongli and Wang, Zhaoyu},
  journal={IEEE Transactions on Smart Grid},
  volume={12},
  number={5},
  pages={3835--3846},
  year={2021},
  publisher={IEEE}
}

@misc{OEDI_Dataset_153,title = {Commercial and Residential Hourly Load Profiles for all TMY3 Locations in the United States},author = {Ong, Sean and Clark, Nathan},abstractNote = {Note: This dataset has been superseded by the dataset found at "End-Use Load Profiles for the U.S. Building Stock" (submission 4520; linked in the submission resources), which is a comprehensive and validated representation of hourly load profiles in the U.S. commercial and residential building stock. The End-Use Load Profiles project website includes links to data viewers for this new dataset. For documentation of dataset validation, model calibration, and uncertainty quantification, see Wilson et al. (2022). <br><br>These  data were first created around 2012 as a byproduct of various analyses of solar photovoltaics and solar water heating (see references below for are two examples).  This dataset contains several errors and limitations. It is recommended that users of this dataset transition to the updated version of the dataset posted in the resources.  This dataset contains weather data, commercial load profile data, and residential load profile data.<br><br>Weather<br>The Typical Meteorological Year 3 (TMY3) provides one year of hourly data for around 1,000 locations. The TMY weather represents 30-year normals, which are typical weather conditions over a 30-year period.<br><br>Commercial<br>The commercial load profiles included are the 16 ASHRAE 90.1-2004 DOE Commercial Prototype Models simulated in all TMY3 locations, with building insulation levels changing based on ASHRAE 90.1-2004 requirements in each climate zone. The folder names within each resource represent the weather station location of the profiles, whereas the file names represent the building type and the representative city for the ASHRAE climate zone that was used to determine code compliance insulation levels. As indicated by the file names, all building models represent construction that complied with the ASHRAE 90.1-2004 building energy code requirements. No older or newer vintages of buildings are represented.<br><br>Residential <br>The BASE residential load profiles are five EnergyPlus models (one per climate region) representing 2009 IECC construction single-family detached homes simulated in all TMY3 locations. No older or newer vintages of buildings are represented. Each of the five climate regions include only one heating fuel type; electric heating is only found in the Hot-Humid climate. Air conditioning is not found in the Marine climate region.<br><br>One major issue with the residential profiles is that for each of the five climate zones, certain location-specific algorithms from one city were applied to entire climate zones. For example, in the Hot-Humid files, the heating season calculated for Tampa, FL (December 1 - March 31) was unknowingly applied to all other locations in the Hot-Humid zone, which restricts heating operation outside of those days (for example, heating is disabled in Dallas, TX during cold weather in November). This causes the heating energy to be artificially low in colder parts of that climate zone, and conversely the cooling season restriction leads to artificially low cooling energy use in hotter parts of each climate zone. Additionally, the ground temperatures for the representative city were used across the entire climate zone. This affects water heating energy use (because inlet cold water temperature depends on ground temperature) and heating/cooling energy use (because of ground heat transfer through foundation walls and floors). Representative cities were Tampa, FL (Hot-Humid), El Paso, TX (Mixed-Dry/Hot-Dry), Memphis, TN (Mixed-Humid), Arcata, CA (Marine), and Billings, MT (Cold/Very-Cold).<br><br>The residential dataset includes a HIGH building load profile that was intended to provide a rough approximation of older home vintages, but it combines poor thermal insulation with larger house size, tighter thermostat setpoints, and less efficient HVAC equipment. Conversely, the LOW building combines excellent thermal insulation with smaller house size, wider thermostat setpoints, and more efficient HVAC equipment. However, it is not known how well these HIGH and LOW permutations represent the range of energy use in the housing stock.<br> <br>Note that on July 2nd, 2013, the Residential High and Low load files were updated from 366 days in a year for leap years to the more general 365 days in a normal year. The archived residential load data is included from prior to this date.},url = {https://data.openei.org/submissions/153},year = {2014},howpublished = {Open Energy Data Initiative (OEDI), National Renewable Energy Laboratory, https://doi.org/10.25984/1788456},note = {Accessed: 2025-06-12},doi = {10.25984/1788456}}

@techreport{doe2025,
  author      = {{U.S. Department of Energy}},
  title       = {Department of Energy Releases Report Evaluating U.S. Grid Reliability and Security},
  institution = {U.S. Dept. of Energy},
  type        = {Tech. Rep.},
  year        = {2025},
  note        = {[Online]. Available: \url{https://www.energy.gov/articles/department-energy-releases-report-evaluating-us-grid-reliability-and-security}}
}

 




\vfill

\end{document}